\documentclass[acmsmall,nonacm]{acmart}
\AtBeginDocument{%
  }

\acmDOI{10.1145/3757686}
\thanks{© 2025 ACM. This is the author's version of the work.
    It has been accepted for publication in Proceedings of the ACM on Human-Computer Interaction (PACMHCI),
    Volume 9, Number 7 (CSCW), November 2025 Issue.
    The Version of Record is available at: \url{https://doi.org/10.1145/3757686}}

\begin{document}

\title{Understanding Collaboration between Professional Designers and Decision-making AI: A Case Study in the Workplace
}
\renewcommand{\shorttitle}{Understanding Collaboration between Professional Designers and Decision-making AI}


\author{Nami Ogawa}
\email{nami_ogawa_pub@cyberagent.co.jp}
\orcid{0000-0002-5361-8934}
\affiliation{%
  \institution{CyberAgent}
  \city{Tokyo}
  \country{Japan}
}

\author{Yuki Okafuji}
\email{okafuji_yuki_xd@cyberagent.co.jp}
\orcid{0000-0002-9547-3681}
\affiliation{%
  \institution{CyberAgent}
  \city{Tokyo}
  \country{Japan}
}

\author{Yuji Hatada}
\email{hatada@cyber.t.u-tokyo.ac.jp}
\orcid{0000-0002-1202-8559}
\affiliation{%
  \institution{The University of Tokyo}
  \city{Tokyo}
  \country{Japan}
}

\author{Jun Baba}
\email{baba_jun@cyberagent.co.jp}
\orcid{0000-0003-0680-5021}
\affiliation{%
  \institution{CyberAgent}
  \city{Tokyo}
  \country{Japan}
}


\begin{teaserfigure}
    \centering
    \includegraphics[width=\textwidth]{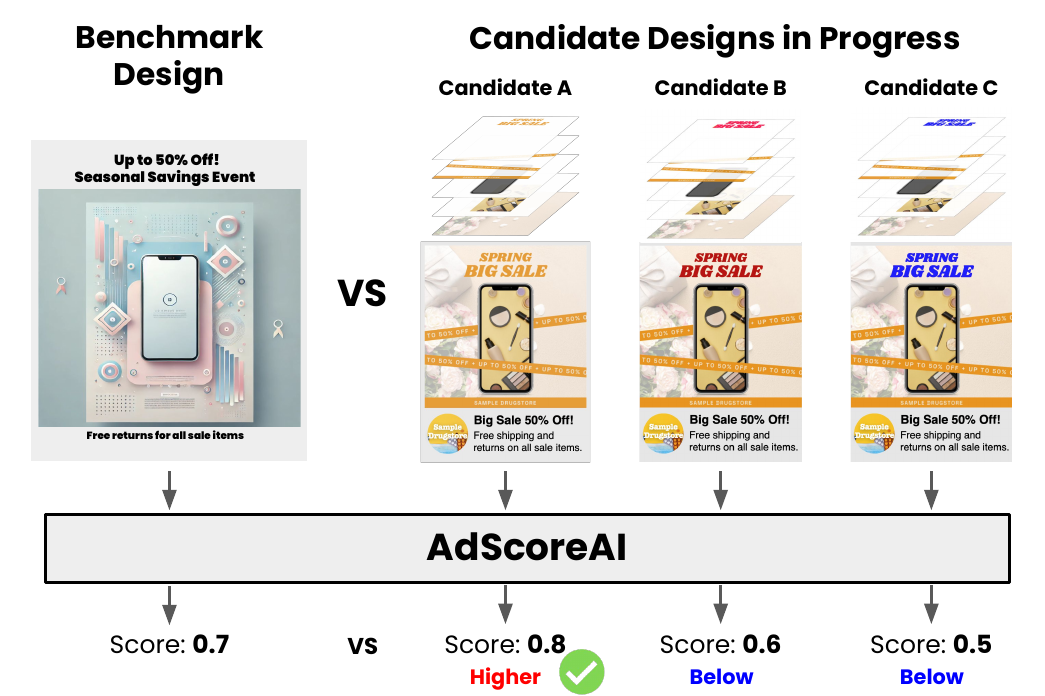}
    \caption{Overview of Design Workflow with AdScoreAI. Designers create advertising designs based on the score feedback from AdScoreAI, which indicates the predictive performance of an input design.}
    \label{fig:teaser}
\end{teaserfigure}

\begin{abstract}
The rapid development of artificial intelligence (AI) has fundamentally transformed creative work practices in the design industry. Existing studies have identified both opportunities and challenges for creative practitioners in their collaboration with generative AI and explored ways to facilitate effective human-AI co-creation. However, there is still a limited understanding of designers' collaboration with AI that supports creative processes distinct from generative AI. To address these gaps, this study focuses on understanding designers' collaboration with decision-making AI, which supports the convergence process in the creative workflow, as opposed to the divergent process supported by generative AI. Specifically, we conducted a case study at an online advertising design company to explore how professional graphic designers at the company perceive the impact of decision-making AI on their creative work practices. The case company incorporated an AI system that predicts the effectiveness of advertising design into the design workflow as a decision-making support tool. Findings from interviews with 12 designers identified how designers trust and rely on AI, its perceived benefits and challenges, and their strategies for navigating the challenges. Based on the findings, we discuss design recommendations for integrating decision-making AI into the creative design workflow.
\end{abstract}



\maketitle

\section{Introduction}
The rapid development of artificial intelligence (AI) has fundamentally transformed the way the creative process works in the design industry.
As AI systems are increasingly being utilized and integrated into creative workflows, it is crucial to understand these implications for practitioners.
Unlike laboratory experiments, the introduction of AI systems in the workplace can profoundly influence designers' creative workflows, compelling them to use AI regardless of their personal preferences or trust in the technology. Such mandated reliance on AI raises important questions about how human designers can adapt to the AI's role in the creative process and how they can effectively collaborate with these systems in a compulsory use setting.

While various forms of AI, including generative AI, have been employed in the creative domain, this study focuses on how designers interact with decision-making AI, which assists or automates the human process of making choices. Decision-making AI typically provides recommendations or decisions to users by analyzing data.
A growing body of research on decision-making AI has underscored the importance of users' appropriate trust in and reliance on AI for practical applications \cite{Kahr2024-bf, Schemmer2023-xu, Yang2020-yl, Adadi2018-kb, Lai2019-ti}, but it has so far focused primarily on domains outside of creative practice.
Decision-making in the creative process, however, has distinctive characteristics: the design process iteratively alternates between convergent thinking, i.e., narrowing ideas through decision-making, and divergent thinking, i.e., exploring possibilities.
Thus, designers must continually make decisions, that is, evaluate options and select promising ideas to refine or finalize, even without a clear goal.
Consequently, in the creative domain, decision-making AI has the potential to support convergent thinking, where designers evaluate, select, and refine a small set of ideas. Conversely, generative AI, which has recently attracted growing attention among both practitioners and researchers, is better suited to support divergent thinking, where designers are encouraged to generate diverse design options while avoiding design fixation \cite{Yoo2024-ms, Choi2024-zc}.

So far, studies on understanding collaboration between creative practitioners and AI have focused on how they use, adopt, and appropriate AI in various creative domains (e.g., sound designers \cite{Kamath2024-mv}, software engineers \cite{Weisz2021-xx, Inie2023-xm, Palani2024-dc}, industrial designers \cite{Uusitalo2024-nx}, science-fiction writers \cite{Palani2024-dc}, UI/UX designers \cite{Uusitalo2024-nx, Palani2024-dc, Inie2023-xm}, and professionals in game industry \cite{Inie2023-xm, Vimpari2023-bo}), but mostly limited to generative AI.
These investigations reveal not only the benefits of AI integration but also the challenges faced by creative practitioners in adapting to changes in their workflows, highlighting the need for further research to understand and foster effective collaboration between humans and AI.
However, generative AI does not support the entire creative workflow of designers and is not the only type of AI.
While recent studies have proposed AI systems to support both divergent and convergent thinking in fashion design \cite{Davis2024-rm, Jeon2021-wu}, a significant gap remains in understanding the socio-technical dynamics that emerge from integrating AI-driven evaluation mechanisms into creative practices, which have the potential to fundamentally reshape decision-making in the design process.
Addressing this gap is essential to inform the design of decision-making AI to support effective human-AI co-creation, and to enrich CSCW discourse on socio-technical system design and the transformation of work.

Therefore, our study aims to fill this gap by examining how the introduction of decision-making AI in the workplace has affected professional designers from their perspective. Specifically, we investigated the following three research questions.
\begin{quote}
RQ1: How do graphic designers trust and rely on AI that offers predictive evaluation in decision-making? What shapes their trust and reliance?

RQ2: What are the benefits and challenges of using decision-making AI for graphic designers in their work?

RQ3: How have they dealt with the challenges?
\end{quote}

To this end, we conducted a case study of an online advertising design company. The case company has used a decision-making AI called \textit{AdScoreAI} (pseudonymous name), which evaluates design using scores, for more than four years to effectively design advertising graphics.
Online advertising was among the first creative fields to use AI for quantitatively evaluating the `quality' of design based on effectiveness.
This is because it has the unique feature that performance can be quantitatively measured through indicators such as click-through rates (CTR) and conversion rate (CVR), leading to early development of models to predict such indicators from images \cite{Cheng2012-qa, Chen2016-kv}.
This context made the company a rare and pioneering example of incorporating a performance prediction model as a decision-making AI in the creative design workflow in the workplace, where the designers create and refine their graphic designs based on the evaluation scores predicted by AI.
By examining long-term designer-AI collaboration, this study provides empirical insight into the evolving human-AI co-creative practice, focusing on decision-making.

Despite the existence of such pioneering examples in the online advertising industry, evaluating quality in other creative design areas remains challenging by nature since the creative field is one of the fields where human intuition and experience are most highly valued.
However, some recent studies have proposed methods to evaluate the design quality of the user interface (UI) and suggested applications that use a score-based evaluation of design quality \cite{Wu2024-jn, Duan2024-zb}, indicating that the creative industry will become increasingly data-driven and incorporate AI in the future.
Such rapid advances in AI technology that can evaluate design suggest the imminent importance of understanding how evaluative AI affects professional designers' decision-making in creative workflows.
To maximize AI's long-term benefits while mitigating potential risks, it is essential to understand its impact on users and anticipate changes in work practices.

In fact, work automation using decision-making AI can sometimes threaten the actual users, i.e., the workers. Previous research has shown that data scientists had mixed perceptions about the automation of work using AI, seeing benefits in speeding up the process but expressing concerns about automating their jobs \cite{Wang2019-wp}.
Additionally, introducing tools to deliver efficiency and support data-driven decision-making to blue-collar workers removed meaning from work and displaced job satisfaction by changing the nature of work and skill requirements, and some workers even developed a shadow system (i.e., informal workaround \cite{Sheehan2023-jw}).
Moreover, in the area of loan consulting in the banking industry, AI has completely substituted for employees' tasks, responsibilities, and decision-making, affecting the identity of their professional roles \cite{Strich2021-fz}.
At the same time, the introduction of generative AI is bringing about various changes in the work of professionals in the creative industry \cite{Palani2024-dc, Inie2023-xm}, from threat-oriented negative appraisals to identification of empowerment opportunities \cite{Uusitalo2024-nx}, necessitating individuals and organizations to recalibrate their work practices \cite{Jaffe2024-ep}.

In this context, this study extends CSCW literature on situated work practices (i.e., context-specific particularities of work practice) and technological support for workers by examining how decision-making AI transforms creative work practices.
Previous research within the CSCW domain has identified and examined changes in work practices across various fields (e.g., healthcare \cite{Sun2023-jl, Su2022-rd}, manufacturing \cite{Sheehan2023-jw}, aviation \cite{Huber2020-td}, IT infrastructure \cite{Wolf2019-tg}) and emerging work practices (e.g., food delivery work \cite{Kusk2022-dj}, crowd workers \cite{Williams2019-co}) due to the evolution of technology. These understandings facilitate the development of technologies and systems to better align with users' actual needs and workflows.
From this standpoint, AdScoreAI serves as a pioneering case study for understanding the impact of incorporating decision-making AI into creative workflow in the workplace through the lens of a professional designer working in the design industry.

In this study, we interviewed 12 professional graphic designers at the company, alongside three managers who are in charge of different areas (i.e., product, operations, and engineering), exploring both the challenges and opportunities of the introduction of decision-making AI in creative fields with a focus on trust and reliance.
Our findings highlight the complex dynamics of integrating decision-making AI into creative workflows. While decision-making AI tools can enhance designers' efficiency and objectively ensure design quality, they also introduce new challenges of changing workflows and conflicts with designers' subjective judgments and norms. 
Design implications for the future introduction of decision-making AI into the creative workplace include: effective collaboration between designers and AI requires clear communication of AI capabilities, implementation of XAI features, and facilitation of task delegation.
Our study contributes to empirical research on workers' perspectives and situated practices with the introduction of decision-making AI in creative fields, and to practitioners by proposing design recommendations.

\section{Related Work}
Given the unique combination of decision-making AI and the creative design field, we first review the use of decision-making AI across various domains in the first half of this section. We introduce the crucial concept of appropriate trust and reliance for effective human-AI collaboration and then highlight the gaps between research and practice in incorporating AI into professional decision-making. In the latter half, we review previous studies on AI in the creative design field.

\subsection{Human-AI Collaborative Decision Making}
\subsubsection{Towards Effective Collaboration: Appropriate Trust and Reliance}
Decision-making AI is used to support human decision-making by providing data-driven insights, predictions, recommendations, or direct solutions to problems.
As these systems are expected to analyze vast amounts of data to identify patterns and trends or evaluate alternatives, it has been studied primarily in domains with significant data availability and a need for objective, empirical decision-making, such as finance, medicine, law, and sports.
In these domains, research has focused not only on developing highly accurate AI but also on enhancing effective collaboration between humans and AI.

For effective collaboration between humans and AI in decision-making, appropriate trust and reliance have been considered critical and are becoming a focus of research \cite{Kahr2024-bf, Schemmer2023-xu, Yang2020-yl, Adadi2018-kb, Lai2019-ti}.
`Appropriate trust' involves users accurately recognizing AI's capabilities, while `appropriate reliance' means users appropriately incorporating AI's outputs, that is, to discern when to rely on AI advice and when to rely on their own judgment \cite{Schemmer2023-xu} or to [not] follow an [in]correct recommendation \cite{Yang2020-yl}. This concept emphasizes that users should not blindly follow AI recommendations but should be empowered to make informed decisions based on the situation, contrasting with the research direction of maximizing AI reliance.
Considering AI is not always perfect, appropriate reliance is considered important to maximize the complementary team performance between humans and AI \cite{Schemmer2023-xu}.

While trust and reliance have been defined in various ways, the consensus is that they are closely related but distinct concepts. Trust is a cognitive state that reflects a user's belief in the AI's capabilities, while reliance is the behavioral manifestation of this trust \cite{Dzindolet2003-ne}, where a user acts based on the AI's advice, measured by the extent to which the AI's advice influences a human's decision-making \cite{Kahr2024-bf, Scharowski2022-zo}.
Research indicates that trust generally increases reliance on AI; however, reliance can occur even in the absence of trust if the user perceives the AI's advice as the best available option \cite{Lee2004-qg}. Conversely, a user might trust an AI system but choose not to rely on its advice in specific situations due to other influencing factors, such as perceived risk or self-confidence \cite{Schemmer2023-xu}.

Several factors, including system, human, and interaction factors, influence trust and reliance on AI systems in decision-making.
System factors include transparent explanations regarding output \cite{Lai2019-ti, Wang2023-ym, Yang2020-yl, Adadi2018-kb}, tutorials about the limitations and capabilities \cite{Chiang2022-hv, Lai2020-es}, and continuous feedback on the performance of the systems \cite{Lu2021-uo}.
In particular, explainable AI (XAI) has been considered important, which aims to improve transparency and interpretability in AI systems, ensuring that users can grasp the rationale behind AI-generated recommendations and predictions \cite{Abdul2018-zc}.
On the other hand, human factors include AI literacy \cite{Chiang2022-hv}, expertise \cite{Logg2019-bd}, and self-confidence \cite{Lewandowsky2000-lc} of human decision-makers. 
Interaction factors include cognitive load \cite{You2022-ah}, risk perception \cite{Green2019-ed}, and other factors that change over time throughout the interaction with the system \cite{Glikson2020-kw, Kahr2024-bf} such as first impressions \cite{Tolmeijer2021-ho}, updates of expectations \cite{Parasuraman1997-un}, experience of error \cite{Rieger2022-ri}, and perceived accuracy \cite{Kahr2023-zs}.

\subsubsection{Research-practice Gaps in Incorporating AI into Professional Decision-Making}
In some industries, decision-making AI has already been introduced in real-world settings and significantly impacted professional workers in the workplace.
Human-AI decision-making systems often take the form of algorithm-in-the-loop, where AI provides recommendations while humans remain the final decision-makers \cite{Green2019-ed}. This collaborative approach is particularly beneficial in high-stakes domains where human judgment is crucial, such as judicial decisions \cite{Hayashi2017-tf}, medical diagnoses \cite{Cai2019-mh}, and school admission decisions \cite{Zhang2023-wx}.
When AI completely substitutes human decision-making, it can impact professional workers' roles and identities. The case study by Strich et al. \cite{Strich2021-fz} provides examples of the introduction of fully automated decision-making AI in the banking industry. The study shows how it has completely replaced the work traditionally performed by loan consultants, reshaping the roles and consequently impacting their professional identity.

In contrast, in a research context, Lai et al. \cite{Lai2023-tf} identified in their survey paper on empirical studies on human-AI decision-making that most existing studies adopt decision-making tasks in high-stakes domains such as medicine, healthcare, finance, criminal justice, and hiring. They point out that experts are rarely included as participants in existing studies, although such decision-making tasks require domain expertise. This highlights a misalignment between study designs and real-world AI applications.
In fact, the levels of expertise affect decision-making behaviors; perceptions of self-capabilities and expertise may moderate the beliefs regarding AI's abilities to perform a specific task \cite{Glikson2020-kw}.
Novices often depend heavily on AI advice due to their lack of experience and domain knowledge, while experts tend to dismiss advice, relying more on their own judgments, even though ignoring it may decrease performance \cite{Logg2019-bd}. 
Additionally, as being an expert means having a current routine, professionals show difficulty in incorporating AI into their routines \cite{Muijlwijk2024-ph, Khairat2018-wp}. 
Such gaps between research and practice and the significant effect of expertise on human-AI decision-making emphasize the need for studies that explore how professionals integrate AI into their decision-making processes within their actual workflows, which motivated our case study in creative design, an underexplored area in existing research.
Specifically, the growing interest in the field of appropriate trust and reliance in literature motivated us to understand how, if at all, professional graphic designers place their trust and reliance on AI.

\subsection{Human-AI Collaboration in Creative Design Fields}
\subsubsection{From Creativity Support Tools to Human-AI Co-Creation}
In the field of human-computer interaction (HCI), Creativity Support Tools (CSTs) have been developed to enhance human creative design processes \cite{Shneiderman2007-qw, Frich2019-pq}, such as ideation \cite{Wang2010-ju}, realization \cite{Davis2015-kg}, and evaluation \cite{Singh2011-bj}. 
With the advancement of AI, the focus of CSTs has increasingly shifted to fostering collaboration between humans and AI, which has since drawn attention in HCI to understanding and facilitating effective interactions.

While human-AI co-creation, i.e., human-AI collaboration in creative tasks, is increasingly becoming common in various domains (e.g., AI that promotes idea generation through group brainstorming \cite{He2024-ie} and journalists utilizing AI in reframing narratives for social media \cite{Wang2024-sz}), designers are one of the most prominent groups that has been radically impacted by the progress of AI.
Researchers have proposed and developed AI tools that can assist designers in discovering, visualizing, creating, and testing (see \cite{Shi2023-uk} for a review).
Research highlights that AI's role in design is evolving, with AI becoming more and more integrated into design workflows.

However, studies suggest that existing AI-supported design tools only work in oversimplified scenarios and are rarely used by industry practitioners \cite{Lu2022-nd}, indicating a need for HCI researchers to bridge the research-practice gap between research findings and real-world applications by investigating how these research results can be applied in real-life practices instead of lab settings \cite{Shi2023-uk}.
Such concerns led researchers to investigate the interaction between humans and AI in the collaborative design process. 
Existing studies have explored how generative AI has been used by creative practitioners, such as sound design practitioners \cite{Kamath2024-mv}, software engineers in code translation \cite{Weisz2021-xx}, user experience (UX) and industrial designers \cite{Uusitalo2024-nx}, creative practitioners in various domains during their real-world projects \cite{Palani2024-dc, Inie2023-xm}, and game professionals in video-game industry \cite{Vimpari2023-bo}.
For instance, Han et al. \cite{Han2024-ff} explored the challenges and strategies human dyads employ to effectively collaborate with generative AI by co-prompting in creating stage designs together and how individuals perceive the role of generative AI. 
Additionally, Ko et al. \cite{Ko2023-dp} studied how visual artists would adopt generative AI to support their creative works and identified three roles of generative AI: automating the creation process (i.e., automation), expanding their ideas (i.e., exploration), and facilitating or arbitrating in communication (i.e., mediation). 
Nonetheless, although there is extensive research on understanding the opportunities and challenges of creative practitioners using generative AI, it does not entirely cover the complex creative processes of professional designers.

\subsubsection{Support of AI for Convergent Thinking in Design Ideation}
The creative process requires designers to explore different options, recreate ideas, and refine solutions; the design process requires repeated evaluation.
In fact, the design is a fundamentally iterative and non-linear process with an open-ended goal \cite{Palani2024-dc, Zhou2024-fd}.
In particular, the ideation process, which is the most commonly supported process by CSTs \cite{Frich2019-pq}, requires both divergent thinking, which involves creating multiple design options, and convergent thinking, which involves selecting and refining a few ideas.
CSTs, especially leveraging generative AI, have succeeded in helping designers with divergent thinking by helping designers overcome design fixation \cite{Yoo2024-ms, Jeon2021-wu, Choi2024-zc}. However, although some research has presented tools for both divergent and convergent thinking in fashion design \cite{Davis2024-rm, Jeon2021-wu}, a significant gap remains in proposing and understanding how designers can utilize AI tools to foster convergent thinking.

In contrast to divergent thinking, convergent thinking involves decision-making, where designers evaluate various options to find the best ideas.
Thus, technology that effectively supports convergent thinking is one that can evaluate design.
Although defining and measuring `quality' in design is challenging by nature, several studies developed models to evaluate design qualities including aesthetics of logos \cite{Zhang2017-in} and car designs \cite{Pan2017-iv}, novelty of graphic design \cite{Wachs2018-ou}, personality (e.g., futuristic, romantic) of posters \cite{Zhao2018-zc}, and design quality (e.g., usability and compliance with design guideline) of UI \cite{Wu2024-jn, Duan2024-zb}.

Applications in UI design areas have been proposed to automatically evaluate design quality using a numerical score and give natural language feedback \cite{Wu2024-jn, Duan2024-zb}, aiming to improve the design. Additionally, Kong et al. \cite{Kong2023-gj} developed a model that can evaluate graphic design considering human aesthetic preferences, allowing the proposed system to automatically suggest an aesthetically improved version of user input for presentation slides.
However, there is still a lack of understanding of how professional designers can adopt such evaluative AI that supports decision-making in their creative workflows and how it impacts them. 

\subsubsection{Graphic Design and AI: From Automatic Generation to Collaboration}
Design encompasses the practical application of creative ideas to address specific problems or opportunities. In fact, while designers and artists are both creative, artists focus on self-expression, whereas designers attempt to find solutions to the needs of targeted users \cite{Shi2023-ua}. Among various areas of design, such as graphic, UI/UX, industrial, and fashion, the most common area for AI research is graphic design \cite{Shi2023-ua}, which combines visual elements (e.g., images, symbols, illustrations, and texts) to communicate messages effectively, often requiring a balance between aesthetics and functionality. Graphic design is not only a common subject in research but also common in our daily lives in various forms, such as posters, banners, presentation slides, and magazines. 

Graphic design involves numerous decisions from the vast array of design choices regarding factors such as font, color, image, and layout, which requires significant time and effort to adjust visual attributes to make them aesthetically appealing and functionally effective \cite{Zhao2018-zc, Guo2021-ky}. This process consists of two main stages: the creative process, where an initial design is formed, and the refinement process, which involves trial-and-error adjustments to enhance the design \cite{Kong2023-gj}. 

Existing studies in graphic design areas have developed and leveraged AI from automatic generation to systems that support design workflows.
For example, design generation is a topic that has gained significant attention, where technologies are developed to automatically create graphical materials such as fonts \cite{Lian2019-yu}, icons \cite{Zhao2020-rz}, and layouts (i.e., arrangement of elements) \cite{Zheng2019-fo, O-Donovan2014-gn}.
Additionally, researchers have developed systems that support specific processes and tasks in graphic design workflows. For example, an interactive language-based interface has been developed for adjusting the photo color \cite{Wang2023-tr}. Kong et al. \cite{Kong2023-gj} proposed a system to support the refinement of a presentation slide design by suggesting a refined version with improved aesthetic quality from the existing design.
Choi et al. \cite{Choi2024-zc} proposed a system that utilizes generative AI to support the divergent ideation process for graphic designers by automating the reference recombination process. Zhou et al. \cite{Zhou2024-fd} proposed OptiMuze, which works as a co-design agent that fosters the exploration and reflection processes of graphic designers.

In HCI areas, only a few studies have focused on advertising design. Vinci \cite{Guo2021-ky} is a system designed to automatically generate posters from user-specified elements, such as product images and text descriptions, considering specific constraints and purposes in advertising posters.
Additionally, similar to the concept behind AdScoreAI, Sun et al. \cite{Sun2018-qn} attempted to enhance television advertising creation by constructing a model to predict the effectiveness of advertisements based on a dataset that continuously measured viewers' emotional responses and viewing interest by facial and eye tracking data.
In summary, as CSTs evolve from simple tools to AI-integrated design systems, it is essential for professional designers to adopt a more collaborative approach, leading to a change in their traditional design workflow.
This study explores how graphic design professionals in the online advertising industry perceive the benefits and challenges of evaluative AI as a decision-making support tool in their design workflow.

To summarize the literature review, the introduction of AI tools into the creative design workflow focuses on improving creativity and efficiency, with generative AI supporting divergent idea generation and execution, while there is a lack of understanding of how evaluative AI supports designers' decision-making processes.
Additionally, although existing research highlights AI's collaborative potential, there is limited understanding of professional designers' collaboration practices with decision-making AI tools in the workplace, where long-term collaboration is necessary. As trust and reliance change over time depending on various factors, it is essential to understand the process of long-term acceptance and trust.
Thus, we investigated how designers in the online advertising industry, who have been using decision-making AI in their work, perceive and experience the AI and how it affects their creative practices.

\section{Case Description: AdScoreAI}
This section provides contextual information about the case of AdScoreAI, a decision-making AI used by the online advertising design company we interviewed. In this company, graphic designers create advertising graphics while interacting with AdScoreAI throughout the entire design process. To our knowledge, the attempt to thoroughly integrate a predictive AI model into the design workflow is unique to the subject in our case study. However, the aim of this paper is not to claim its novelty. Rather, we believe that as AI models that measure, evaluate, and predict aesthetic and functional qualities continue to develop in any creative design field (e.g., UI design evaluation \cite{Wu2024-jn, Duan2024-zb}), designers' workflows will increasingly incorporate such models. For this reason, we have chosen this case as a pioneering example of integrating a predictive AI model into creative design workflows as a decision-making AI to understand the interaction between designers and the decision-making AI.

The descriptions in this section are based on information gained in interviews with the managers and observations from the on-site visit (see \autoref{method:interviews-managers} for the protocol).
However, \autoref{subsec: industrial} is also corroborated by a literature survey.

\subsection{Industrial Context: Online Advertising Industry}\label{subsec: industrial}
Despite being a creative field, advertising design primarily aims to promote products or services, making it highly functional. This distinguishes it from other forms of graphic design that focus more on aesthetics.
Moreover, online advertising uniquely allows for the objective and quantitative measurement of its performance. Metrics such as CTR and CVR are commonly used to gauge the effectiveness of online advertisements. Consequently, designers in the online advertising industry are incentivized to create advertisements that achieve high performance based on these metrics, regardless of the use of AI.

These characteristics of online advertising, namely that the performance can be easily defined and measured and that the design is highly performance-oriented, have led the industry to be one of the first areas of creative design to develop machine learning models to predict the objective quality of design.
In a research context, algorithms and models have been developed to predict the CTR of an advertisement based on its content (i.e., texts, images, and videos) and various contextual information (e.g., user features, advertiser and publisher features, and historical interaction data) \cite{Zhou2016-xl, Richardson2007-lk, Azimi2012-rh, Yang2019-mr}.

In sum, it is common practice in the online advertising industry for designers to create designs while referring to objective performance metrics such as CTR and CVR, and researchers have developed models that can predict such metrics from input design.
Nonetheless, incorporating such predictive models into the designer's workflow and substituting the designer's decision-making is unique to our case.

\subsection{Organizational Context}\label{subsec: organization}
The organizational subject of our case study was an online advertising design company that specializes in creating online advertising graphic designs, including images and videos.
During the period of our research, between July and October 2024, the company had approximately 200 employees, with around 90\

The process a graphic designer goes through when working on a project of a single advertising design is as follows:
\begin{enumerate}
    \item The designer receives a request from a salesperson at the parent company and reviews the client's requests and requirements in consultation with the salesperson.
    \item The designer creates the design using the AdScoreAI interface, considering it complete when the final output score surpasses the standard threshold and the designer determines that quality standards are met.
    \item The work is submitted to the sales department. In principle, no other designer reviews it before submission.
    \item If the client requests revisions through the sales department, the process returns to step 2.
\end{enumerate}

One designer is in charge of each design. In other words, it is uncommon for multiple designers to collaborate on a single design.
Designers typically submit around nine completed designs per day. 
This means that they spend around 50 minutes on each piece of work, but the design time varies from around five minutes to around one hour.

There is always a trade-off between time and quantity in creative design tasks, including general advertising design. In other words, while design time does not necessarily directly correlate to quality, there is a balance to be struck between the quality improvement that can be expected by spending more time on each design and the quantity of designs that can be increased. From this perspective, this company is positioned within the industry as one that places importance on the number of designs it produces. This decision is a strategy that has been adapted to the online advertising distribution system in recent years. Since the advertising distribution platform uses algorithms to deliver the most relevant advertisements to target users, increasing design variation is expected to improve advertisement effectiveness. 
This policy existed even before AdScoreAI was introduced in 2020.
Since the initial introduction, the AI model algorithms, data used for learning, design support tools using the model, and the design workflow using the tools have all continued to be updated.

\subsection{AdScoreAI as a Predictive AI Model}\label{subsec: model}
AdScoreAI is a machine-learning-based AI model that can evaluate the quality of an advertising design. AdScoreAI and the design support tool that incorporates AdScoreAI are both proprietary in-house systems developed by the parent company.
In this subsection, we explain the underlying AI model. AdScoreAI takes multimodal media (e.g., images, videos, text) and metadata (e.g., distribution settings and targeting configurations) as input. Using proprietary metrics based on combinations of performance metrics such as CTR and CVR, it predicts expected performance if the advertisement is distributed and outputs a single numerical score. A higher score means that the input media is more promising as an advertisement. This score is referred to as the `score feedback' or simply `score' hereafter.

The model architecture consists of three encoders, each handling a specific input modality: a Convolutional Neural Network (CNN) for an image or video, a Long Short-Term Memory (LSTM) network for texts, and a Multi-Layer Perceptron (MLP) for tabular data representing the distribution settings (e.g., delivery methods and placement) and targeting configurations (e.g., audience demographics).
The model is regularly trained using advertising distribution data, including advertisements and their performance, provided by the parent company that developed it. As of 2024, the model has been trained using approximately 100,000 advertisement data sets. In the early stages, however, the model architecture was different from the current one, and the amount of training data was much smaller. Since its introduction, the accuracy of the scores has been improving through model updates and training.

\subsection{AdScoreAI-integrated Design Workflow}\label{subsec: workflow}
When creating advertising designs, the graphic designers at the company must use a browser-based interface that integrates AdScoreAI. When a designer uploads media (i.e., images or videos) to this interface, it internally calls AdScoreAI to display the evaluation score. 
Designers can upload multiple media simultaneously and compare their scores. Additionally, this interface features a comparison with the project's benchmark media, which is the advertisement with the highest performance in the project's current distribution. The evaluation score of this benchmark media is displayed alongside the designer's uploads, enabling a comparison by scores.
Although the AdScoreAI model requires metadata regarding the project as well as media input to output a score, the metadata is specific to each project and fixed. Therefore, what designers upload to this interface are the design candidates they are creating. Since the metadata does not usually change during the course of a project, the score simply indicates how well the uploaded media is expected to perform under the given project settings.

The operational rules require designers to create new designs that achieve a higher score than the benchmark score (i.e., the ``baseline''). Designers call it a ``win'' when they exceed the benchmark score. Conversely, if the scores of the uploaded media are below the benchmark, it is termed a ``loss.'' In such cases, designers must refine their designs until the deliverable achieves a win.
It is noteworthy that the designers not only use the score to improve the potential effectiveness of the deliverables but also consistently compare the scores of their designs in progress with the benchmark score during the design workflow. The designers always upload multiple candidates, compare the candidates' scores with the benchmark score, and select the candidate that is superior to the benchmark by score. In other words, the standard by which designers judge whether a design is ``better'' is always based on the scores throughout the design process. 
This unique design workflow is achieved by combining the AdScoreAI model with interface design and operational rules in an inseparable manner.
Both the interface and rules have changed over time since the first introduction. Additionally, the operational rules have detailed variations, depending on specific circumstances such as media type (e.g., an image or video) and distributing platform. However, the basic rules are that the score of the deliverable must outperform the benchmark score, and the candidate with the higher score is considered to be preferable, which is a common standard.

This design workflow significantly differs from conventional workflows that do not utilize AI, which are still common in the creative design field and were previously used in this organization before the introduction of AdScoreAI. In traditional workflows, designers typically start with a rough draft of the final product and gradually refine details until completion, leaving the steps leading up to the final product largely at the discretion of the individual designer. In contrast, the current AI-driven workflow involves a repeated selection process for each step, making the design process inherently bottom-up.
Note that, as the designers never access the AdScoreAI model directly outside of this design support interface that incorporates AdScoreAI, they may not clearly distinguish between the model and the interface, and refer to them both as AdScoreAI.

\section{Method}
Taking AdScoreAI as a case study, we conducted online user interviews with 12 graphic designers who used AdScoreAI for their work, and analyzed the data to gain insights.
Additionally, we interviewed three managers (i.e., product, operations, and engineering managers) to grasp the general idea of AdScoreAI and its usage from an organizational perspective.
Furthermore, there was one on-site visit to the advertising design company prior to the graphic designer interviews, which enabled us to gain an in-depth understanding of AdScoreAI and design workflow and helped us develop tentative research questions.
All stakeholders, including the authors, use Japanese as their native language, and all research processes except for writing papers were conducted in Japanese.

\subsection{Interviews with Graphic Designers}

\subsubsection{Participants}
Interviewees of the user interviews consisted of 12 professional graphic designers.
The interviewees were recruited through the operating manager at the company using maximum variation sampling.
That is, prior to conducting the user interviews, the 12 interviewees were selected to ensure diversity in terms of the type of media they were working with (i.e., image or video), years of service, prior design experience (i.e., whether or not they have a degree in design, and whether or not they have previous work experience of advertising design), and gender.
We determined in advance that if we did not reach theoretical saturation in the user interviews with the 12 interviewees, we would ask for additional interviewees. Nonetheless, no further new insights arose from the interviews, and thus, we finished the data collection when we completed the interview with the 12th interviewee.

The 12 interviewees' average age was 27 (min 24, max 34) years old, and tenure with the company is 2.75 (min 2, max 6) years. They consist of four men and eight women (self-reported). Although we collected demographic information for each interviewee, we only report statistics and do not provide a correspondence table to avoid identification through demographic combinations.

In exchange for not being compensated, they were asked to participate in the interviews as part of their work. After the operating manager had selected the candidates in advance, the first author explained to each participant the research purpose, that participation was voluntary and that there were no disadvantages to not participating, the benefits and disadvantages of participating, and the ethical considerations mentioned below. After obtaining consent, we conducted the interviews. None of the candidates declined to participate.

\subsubsection{Researcher Characteristics, Reflexivity, and Researcher-participant Relationship}
The authors consist of researchers specializing in the HCI field. All of them have no experience in engaging in the creative design business or research. 
Some of the authors, including the first author, are researchers who belong to the same company as the company that develops AdScoreAI, which is the parent company of the design company that the interviewees work for.
Such relationships may have potentially affected the power dynamics between the interviewees and the interviewers (i.e., the first and second authors), although there was no prior acquaintance between the interviewers and interviewees.
Note that the other author is a researcher working at a university, so there was no direct conflict of interest (COI) with the participants. 

Given the hierarchical relationship and COI, the following ethical considerations were made for the interviews with the graphic designers. 
First, we asked the interviewees' managers in advance for their cooperation so that the interviewees would not be disadvantaged in their work. Specifically, we requested that their decision to participate or not and the remarks made during the interviews not influence their evaluation, and that their workload be adjusted so that the interviews would not interfere with their regular work.
We also ensured interviewees would not be blamed for any harms, such as work disruption or information leaks.
We also asked that they not make any deliberate assumptions that could identify individuals from the published materials, and that they not use them for personnel evaluation or labor management.
We explained these points to the managers and obtained consent, then informed the interviewees before each session.
Additionally, we took strict measures to ensure only the authors could access the recordings, and participants were informed accordingly.

\subsubsection{Interview}
The study was approved by the IRB of the lead author's institution.
The user interviews were conducted using the online video conferencing tool (i.e., Zoom) with recording between September and October 2024.
The user interviews were conducted by the first and second authors. The first author was the lead interviewer in all the interviews, and the second author was present for half of the interviews, asking additional questions as an assistant interviewer. For each designer, we conducted one individual interview lasting 90 minutes, including an explanation of the interview and consent to it.
The average interview time, excluding the explanation period, was 1 hour 11 minutes, ranging from a minimum of 1 hour 0 minutes to a maximum of 1 hour 20 minutes.
The interviews were unstructured. Thus, although the questions were in line with the interview guide, the interviews were conducted in exploratory and flexible manners in principle.

The typical questions were as follows: how they use AdScoreAI, how AdScoreAI had affected their work, the benefits and challenges they perceived, and how the perception of AdScoreAI and/or score had changed over time. 
We had a preliminary interest in trust and reliance based on existing scholarship, as represented in RQ1. However, rather than using direct questions such as \textit{``Do you trust [rely on] AdScoreAI?'',} we sought to understand these perspectives through related indirect questions. These included questions such as \textit{``How do you feel about the score?''} and \textit{``How do you perceive it?''} as well as \textit{``How do you incorporate the score into your work?''} and \textit{``How do you use it?''} This approach was adopted due to the subjective nature of trust and reliance, which can vary in definition and meaning among individuals.
Nonetheless, the terms `trust' and `reliance' often came up naturally in the course of the conversation, both from the interviewees and the interviewer. At such times, the interviewers delved deeper into it.
On the other hand, regarding RQ2, we asked direct questions on their perceived benefits, challenges, conflicts, and coping, then delved into the details.

\subsection{Interviews with Managers and On-site Visit}\label{method:interviews-managers}
Additionally, we interviewed managers to grasp the overview of AdScoreAI and its usage from an organizational and technical perspective.
Between May and October 2024, interviews with the managers were conducted repeatedly as necessary.
Most of these interviews were conducted before the user interviews with the graphic designers, so that we could conduct the user interviews smoothly by understanding background information beforehand. This included grasping the industry structure, designer tasks, and terminology used in the organization and industry.
However, some interviews were held later to verify facts and support the interpretation from organizational and technical perspectives. 
We interviewed three managers: an operations manager at the advertising design company, which designers work for, and a product manager and an engineering manager at the company that develops AdScoreAI, which is the parent company of the advertising design company.
Specifically, the interviews with the product manager focused on their intended goals in implementing and introducing the AI system.
The interviews with the operations manager focused on the organizational operational rules, their decision-making process, and their intentions.
The interviews with these two managers helped us grasp the case and context described in \autoref{subsec: organization} and \ref{subsec: workflow}. In other words, the descriptions in these subsections are based on what we learned from these two managers.
The interviews with the engineering manager focused on the technical details of the model of AdScoreAI, including its functions and theoretical limitations, which helped us understand and describe the content in \autoref{subsec: model}.
The authors directly contacted the managers, explained the purpose of the study, and asked them to be interviewed as part of their work.

Besides the interviews, there was one on-site visit to the advertising design company in July 2024, prior to the designer interviews.
All authors made observations during the on-site visit, which enabled us to gain an in-depth understanding of AdScoreAI and the design workflow and helped us formulate tentative research questions.
During the on-site visit, the authors observed the designers at work and were given a detailed explanation of the design process from the operations manager and designers.

These interviews with the managers and the on-site visit were all informal sessions and were not fully recorded by video or audio. Thus, the data from these interviews were not directly included in the data analysis that follows. Nonetheless, the authors each kept notes based on the interviews or their observations, and these notes were shared among the authors.
As mentioned above, these interviews with the managers provided us with background information and prior knowledge, which provided useful background that facilitated the user interviews, and made it easier to be familiar with and interpret the data (e.g., understanding the meaning of technical terms and abbreviations, and distinguishing between what they perceive from their perspectives and the technical/operational facts).

\subsection{Data Analysis}
We qualitatively analyzed the collected interview data with graphic designers iteratively, inductively, and reflexively using thematic analysis \cite{Braun2006-pc} as follows. We collaborated on the analysis using Google Spreadsheets.
(1) The second author transcribed the recorded interviews verbatim. 
(2) By reading the transcripts line by line, the first and second authors became familiar with the data.
(3) The first author highlighted the evocative parts and coded each transcript with in-vivo labels, which have summarizing properties, and category labels, which have classification properties indicating the topic.
(4) Each time the data coding for the three interviews was completed by the first author, the first and second authors discussed the labels and themes to refine them and to identify preliminary themes.
During this analysis process, we had multiple discussions with all the authors.
In the discussions, we iterated the following: reviewing the codes, grouping the codes with common points together to form themes, and refining the themes according to the codes until all authors agreed on the themes.
Finally, the themes were grouped according to each research question.

Although most of the authors are affiliated with the company that developed AdScoreAI, as described above, we did not aim to promote AdScoreAI through this paper, and the analysis was conducted from a thoroughly academic perspective in line with the RQs. However, we took care to ensure that we did not quote any statement containing potentially confidential or identifiable information. This editorial censorship was considered in the process of writing and, thus, did not influence the themes identified in the analysis. Additionally, as mentioned above, the authors included one affiliated with independent organizations, playing a role in maintaining neutrality in the entire research process.

\section{Findings}
\subsection{Designers' Trust and Reliance on AdScoreAI}
\subsubsection{Optimistic and Pessimistic Views of Uncertainty}\label{subsubsec: uncertainty}
As AdScoreAI is a machine learning-based prediction model, it is characterized by inherent output uncertainty \cite{Benjamin2021-fo}. Designers' trust depended on how they interpreted such uncertain outputs. 

P8 mentioned, \textit{``I can't say I fully rely on it yet because there's no certainty.''} Similarly, P11 stated, \textit{``I don't trust the score itself because there's no absolute guarantee that a high score will always perform well,''} highlighting that the lack of certainty is a barrier to trust. P12 expressed this uncertainty more specifically by saying, \textit{``Although all deliverables basically exceed the baseline score, not all of them perform better than the benchmark.''}

In contrast, P6 noted, \textit{``I think the actual performance shows the true effect, while the score indicates potential effectiveness.''}
Similarly, P3 stated, \textit{``The score gives an idea of the trend\ldots it's an indicator that guarantees a certain level of effectiveness to some extent,''} acknowledging the lack of certainty but emphasizing a more comprehensive perspective.

\subsubsection{Trust in Objectivity that Humans Lack}
Several designers focused not only on the quantitative aspect of prediction accuracy but also on the quality (i.e., semantics) of the scores. For example, P6 stated, \textit{``I trust the objectivity of AI,''} and P3 commented, \textit{``I believe design cannot be measured solely by data. But that's exactly why I value the score\dots it shows how [the design] stands when viewed through data alone.''} These statements reflect trust in the AI's ability to evaluate from a perspective different from that of humans.
P6 elaborated:
\begin{quote}
\textit{I think machines are less arbitrary [than humans]. They generate scores based on vast amounts of data rather than preferences, which makes them trustworthy. The sales team and I have biases and preconceived notions like ``this is trending now.'' But such assumptions can be disruptive\ldots I think AI is fairer in that sense.} (P6)
\end{quote}

P6 added \textit{``I trust myself less [than AI]''} and P11 also expressed a similar view by saying \textit{``I don't trust my sensitivity,''} suggesting that their self-assessment of their abilities influences their expectations of AI.
Nonetheless, the following quote from P6 indicates that s/he believes that objectivity is necessary in commercial design, implying that P6 trusts AI more than themself in that sense.
\begin{quote}
\textit{For example, if a woman in her 20s is creating advertisements for men in their 60s, I feel her subjective judgment wouldn't make sense. But since I don't have a 60-year-old man to ask, ``What do you think?,'' I would rely on AI.} (P6)
\end{quote}

\subsubsection{Fluctuating Trust}
Designers experienced events that changed their level of trust and reliance on AdScoreAI. These events can be categorized into experiences during the design process and post-production (i.e., evaluation of predictive performance).

Throughout the design workflow, designers repeatedly input the in-progress designs into AdScoreAI and check the scores. Each time, they assess whether the scores seem reasonable, adjusting their trust in AI. For instance, P1 explained the experience of deepening their trust in AI when \textit{``I tried incorporating elements of the benchmark design into the design I was working on, and the score really improved.''} Conversely, P5 expressed complete distrust in AI, stating, \textit{``In a project aimed at women, an image of the Earth exploding received a high score. How on earth can I trust an AI that makes such evaluations?''}

The designers' trust in AdScoreAI can also fluctuate after the design is completed. 
In fact, it has been a common work practice for designers in the online advertising industry to monitor the performance of their advertisements post-delivery, regardless of whether AI is used.
The following comment by P11 illustrates how trust frequently fluctuates depending on the actual performance, leading to the perception of [un]certainty, as discussed in \autoref{subsubsec: uncertainty}:
\begin{quote}
\textit{When the ad performs well, I feel I can trust AI, but when it doesn't, I don't. If the higher-scoring ad performs better, my trust in the score deepens.} (P11)
\end{quote}

\subsubsection{Rules and Trust Shaping Reliance}
Regarding the extent to which the designers followed AI's evaluation (i.e., reliance), designers viewed following the rule of ``exceeding the baseline'' as a good compromise. P9 said, \textit{``My recognition is, as long as the score is above [the baseline], it's just good enough.''} and P3 echoed, \textit{``I feel like it's okay as long as it exceeds [the baseline].''} However, while P9 adhered to the rules because \textit{``there's a rule,''} P3 pointed out the influence of time constraints, stating, \textit{``The real reason is that there's no time. I think, `As long as it exceeds the baseline score, it guarantees some level of effectiveness,' so I accept it.''} Reflecting this pragmatic view, P3 described their trust as, \textit{``It's not like I over-trust it, but I see it just as a means.''}

P9's quote illustrates how a decline in trust in AI undermines reliance on AI over time:
\begin{quote}
\textit{Initially, I believed that the higher the score, the better the performance, so I focused on how much I could exceed the existing score. But as I observed the actual performances, I started to feel that a high score might not matter much. Now, I think it's enough if it just exceeds the baseline. The rules might be a big reason [for my current view].} (P9)
\end{quote}

\subsubsection{Need for XAI}
Several designers mentioned the need for XAI features.
P7 stated that \textit{``if we could understand why this [benchmark] is strong, we could use that [insight to guide our revisions],''} and pointed out that specific feedback is necessary to improve the efficiency of revision work. P10 also stated that \textit{``It would be instructive if [AdScoreAI] could tell us why [my designs] scored poorly,''} and emphasized that in the current situation where feedback is lacking, \textit{``we can only refine on an exploratory basis,''} which causes inefficiency and stress. P5 also expressed dissatisfaction with the AI's outputs, saying that \textit{``the fact that [AdScoreAI] doesn't tell me the reasons [behind the scores] is a huge stress to me\ldots they only reject the design, they don't tell me what is wrong.''}
Furthermore, P5 stated that \textit{``with only AI [without any human designer feedback], I don't feel like I'm growing as an individual [designer],''} and feels that the one-sided score presentation by AI is depriving them of opportunities to gain experience and improve their skills. P9 also stated that \textit{``if the criteria and factors for the score were clearly fed back, I could make it better,''} expressing their expectation that feedback would lead to improved efficiency and opportunities for growth.

\subsection{Benefits and Challenges of Incorporating Decision-making AI into Creative Workflow}

\subsubsection{Double-Edged Sword for Efficiency}
P1 described their work as a \textit{``battle against time,''} stating, \textit{``When time is tight, I feel more frustrated than grateful to AdScoreAI.''} In such time-constrained work environments, designers place great importance on work efficiency, and they perceive AdScoreAI as both beneficial and detrimental in this regard.

\textbf{AI as a Decluttering Partner.}
Designers appreciate that AI shortens the time spent deliberating design choices, thereby improving work efficiency. P12 stated, \textit{``AI chooses for me, so it reduces the time I spend thinking.''} P1 mentioned, \textit{``It puts an end to the endless design time,''} and P2 referred to it as \textit{``a tool that helps reduce the time spent being torn between options.''} P7 noted that not needing to consult others also contributes to efficiency:
\begin{quote}
\textit{When I'm unsure, AI provides the correct answer, which is helpful. I used to frequently ask others, ``What do you think of this material?'' but now I don't need to, and that speeds up my work.} (P7)
\end{quote}

This time-saving aspect also reduces cognitive and mental burdens associated with decision-making. P12 remarked, \textit{``Especially when there are many projects and limited time, it's easier to let AI choose.''} P9 uniquely likened AI to a \textit{``decluttering partner,''} stating, \textit{``I'm really grateful when AdScoreAI helps me organize my thoughts. [\ldots] I might see AdScoreAI as a partner in decluttering. Decluttering involves letting go of things you don't want to discard, which is similar to design. When I have many design ideas like, `I want to do this and that,' AI scores and helps cut things down, so I'm glad to have AdScoreAI.''}

\textbf{Overhead from Unnecessary Fine-tuning.}
All designers except P6 and P11 expressed dissatisfaction with the time-consuming \textit{``adjustment work''} required to achieve a winning score. P1, P3, and P10 felt that this process \textit{``wastes time.''}
As P9 put it, \textit{``I sometimes wonder if it's necessary to spend so much time just to win,''} and P12 added, \textit{``When I spend a lot of time on repeating minor adjustments to finally win, it feels like a waste.''} P12 also admitted that in time-constrained situations, they often adopt \textit{``non-essential strategies''} to win.

This inefficiency directly leads to mental stress under time pressure. P4 mentioned, \textit{``When I keep losing, it becomes stressful because I feel like I'm not making progress.''}
Thus, while AdScoreAI enhances work efficiency by accelerating decision-making, it sometimes reduce efficiency in certain situations, causing dissatisfaction among designers.

\subsubsection{Influence of Incorporating Objective Metrics into Design}
Another major advantage emerged: the score serves as an objective indicator of effectiveness, increasing designers' confidence in their deliverables and simplifying external explanations. However, designers also experienced conflicts.

\textbf{Scores as a Guarantee of Effectiveness.}
Designers found advantages in using the AI score as an objective indicator that guarantees effectiveness. P11 described AdScoreAI as \textit{``providing a basis for confidently delivering work,''} and P1 described it as \textit{``I feel like I'm getting a seal of approval.''} P12 explained this benefit as follows:
\begin{quote}
\textit{My own selection is subjective, but running it through AI adds another layer of validation based on vast data. It feels like saying, ``AI chose it, so it must be good.'' It doesn't mean AI takes full responsibility, but I feel like I'm backed up. I do create with my own hypotheses, but thinking ``AI decided, so it's fine'' gives me confidence and a reason.} (P12)
\end{quote}
P6 shared a specific success story:
\begin{quote}
\textit{When proposing a bold idea, I always tend to get nervous, but if AI scores it highly, I feel confident to present it.} (P6)
\end{quote}
Additionally, P10, who joined the company with no prior experience, appreciated that \textit{``AI compensates for my lack of experience and knowledge in terms of effectiveness.''}

This ``validation by AI'' not only boosts confidence in the deliverables but also serves as objective proof when assuring clients of the effectiveness, as P6 noted, \textit{``AI assures a certain level of effectiveness, so I can confidently tell clients, `It will be effective.' ''}
Additionally, designers appreciated the convenience of not needing to articulate their design intent explicitly as follows: 
\begin{quote}
\textit{When asked about the design intent by the sales team, it's convenient to say it's based on AI results.} (P12)

\textit{When asked, ``Why this image?'' I can simply say, ``Because it has a winning score,'' which serves as a rationale.} (P5)
\end{quote}

However, P12 added, \textit{``AI is doing part of the thinking that designers should be doing,''} and P1 expressed a similar sentiment, stating, \textit{``While it's beneficial to have AI provide reasoning, it can also be a disadvantage for designers if they can't explain their own choices.''} This highlights the designers' concern that short-term gains might lead to long-term disadvantages. Nonetheless, it is important to note that this reduced need to verbalize design intent was a contributing factor in the success story shared by P6.

\textbf{Subjective-Objective Conflict.}
While the assurance of effectiveness provided by AI is beneficial, on the flip side of the benefit, designers often encounter situations where their personal sensitivities conflicts with AI's scores. P8 remarked, \textit{``Sometimes the configurations and color schemes deemed best by AI seem chaotic from a human perspective\ldots our sensibilities differ.''} P7 elaborated on the conflict:
\begin{quote}
\textit{There are times when I don't agree with a high-scoring design. It might be effective, but as a design, it doesn't look good. This conflict is frustrating.} (P7)
\end{quote}

While some designers have found ways to cope with and reconcile this conflict, others continue to struggle. For example, P11 initially felt \textit{``confused''} but now describes the relationships with AI as a \textit{``partnership,''} adding, \textit{``When AI scores something high that I didn't expect, I find it interesting and take it positively.''} In contrast, P5 continues to struggle, saying, \textit{``When AI favors designs where texts are hard to read, it's hard to accept.''} P5 further emphasized that this stress is specific to decision-making AI, stating, \textit{``If a generative AI suggested options that I could choose from, I wouldn't have this stress.''} 

\subsection{Approaches to Dealing with Challenges}
Designers adopted two main strategies to cope with this conflict.
\subsubsection{Assignment of Duties}
One strategy involved understanding how to appropriately place trust in AI.
P6 reflected on past over-reliance on AI, saying, \textit{``Initially, I trusted AI too much and didn't understand the division of roles between AI and creators.''}
P3 remarked, \textit{``[now] I rarely disagree with the scores,''} explaining \textit{``Even if I think differently, I accept [what AdScoreAI says], thinking a high score indicates effectiveness from an objective standpoint.''}
Nonetheless, designers do not blindly follow AI. P11 said, \textit{``I trust the AI's output and align with them, but I still make slight adjustments.''}
Many designers believe that the overall quality of design is ultimately in human hands: 
\begin{quote}
\textit{In the end, it's humans who refine the design.} (P9)

\textit{While AI can raise a certain level of quality, the visual quality depends on the designer's skills.} (P2)
\end{quote}
In contrast, one of the most experienced designers, P5, who has been creating advertisements for the company since before AdScoreAI was introduced, continues to express distrust and struggle to work with AI. While P5 acknowledges AI's role in evaluating effectiveness, P5 believes, \textit{``Design is about making visual adjustments to enhance effectiveness,''} considering both aesthetics and effectiveness as inseparable aspects of design. This contrasts with other designers who separate responsibilities, assigning effectiveness to AI and aesthetics to themselves.

\subsubsection{Reversing Creative Thinking Process}

Some designers pointed out that \textit{``the fundamental process of thinking is completely reversed (P11)''} when using AdScoreAI compared to traditional design workflows. With AdScoreAI, designers tend to \textit{``start without imagining the goal (P11),''} whereas traditional design involves \textit{``creating with the final product in mind (P2)''} and \textit{``working backwards (P5).''} This shift initially caused confusion, as P12 described, \textit{``The direction I wanted to go was completely different from what AI suggested, which was confusing at first.''}

Specifically, designers noted that personal preferences hinder an AI-driven decision-making process. P2 mentioned, \textit{``I used to choose materials based on my preferences, which defeated the purpose of using AI.''} P3 shared, \textit{``At first, I was worried that I wouldn't be able to create the designs I wanted, but now I'm used to it and can create my desired designs with AI.''}
By understanding and adapting to this new dynamic, designers have shown that it is possible to effectively integrate AI into the creative process, demonstrating the potential for harmonious collaboration between human creativity and AI assistance.

\section{Discussion}
The findings from interviews with professional graphic designers who use AdScoreAI, an AI tool that predicts the performance of advertising designs, demonstrate that AdScoreAI supports the designers' decision-making process, which brings them both direct and indirect benefits, but also identifies challenges and psychological conflicts faced by the designers.
In this section, based on the summary of the findings, we discuss future directions for better supporting creative professionals by leveraging decision-making AI.

\subsection{Summary of Findings}
The decision-making AI introduced in the workplace to improve productivity and standardize workflows has been shown to benefit individual designers in their daily tasks. Specifically, it reduces the time and psychological burden associated with the constant decision-making in design and provides objective evidence of effectiveness that offers psychological reassurance, enabling designers to propose their designs with confidence. However, the rigid, score-driven creation process also posed challenges, such as requiring designers to spend what they felt was excessive time tweaking designs to improve scores and causing conflicts when subjective judgments differed from AI-generated scores.

Our findings indicate that the expected role of decision-making AI from designers is more about efficiency and objectivity rather than creativity.
Designers view AI's predictions as fair judgments that are based on a vast amount of data, not influenced by subjective factors. This perspective tends to increase trust in the system, especially among those designers who feel uncertain about their own judgments. Additionally, decision-making AI fosters efficient idea convergence by serving objective metrics for comparison, being trusted as a `decluttering partner.'

\subsection{General Discussion}
Unlike decision-making in healthcare, law, or lab settings with ground truth labels (e.g., \cite{Lai2019-ti, Schemmer2023-xu}), creative design involves open-ended goals. Therefore, designers continually make small decisions throughout the process (e.g., deciding what to add and which to choose, etc.), not just at the final stage (e.g., deciding whether a design is complete). Correspondingly, although AdScoreAI offered a simple single output, it was used in two different ways: first, to objectively compare the quality of design candidates, enhancing the efficiency of making choices; second, to provide objective metrics, ensuring quality. While prior work on decision-making AI focused on its evaluative use, our findings emphasize the importance of its selection-support role, i.e., support for the convergent process, in creative domains.

Additionally, comparing our findings with previous studies on collaboration practice between designers and generative AI, we find both commonalities and differences. Palani and Ramos \cite{Palani2024-dc} identified three benefits of generative AI tools: helping develop creative goals further, streamlining the creative process, and accelerating the generation of alternatives, alongside three challenges: articulating goals, lack of memory across fragmented workflows, and aligning and assessing stochastic model outputs with intents. The benefit of streamlining the creative process aligns with our identified efficiency benefits, but while generative AI automates or supports processes such as initial sketches, mockups, and layout generation \cite{Palani2024-dc}, decision-making AI supports the selection of the most suitable options from multiple candidates. That is, while generative AI generates alternatives, decision-making AI facilitates convergence, offering complementary benefits to designers. Additionally, the benefit of the `Guarantee of Effectiveness,' identified for decision-making AI in our study, is unique and not paralleled in generative AI in prior studies. It is worth noting that, despite the fact that predictions are inherently uncertain and AI cannot ensure definitive outcomes, and even though designers are aware of this fact, they still feel a subjective sense of ``assurance'' of effectiveness as a benefit.
In comparing identified challenges, it is evident that each technology presents unique challenges. Thus, designers derive complementary benefits from generative and decision-making AI but must address and balance the specific challenges each presents.

Moreover, our workplace case, where AI use is mandatory, complements previous studies in lab settings and voluntary use of AI. The pragmatic but distinctive constraints of our case as a practical example of the use of AI in the workplace include strict adherence to AI usage rules and productivity pressures. Specifically, designers improved the scores until they exceeded the baseline, even though they sometimes felt it was `unnecessary fine-tuning.' If using the system was voluntary, this dilemma would not arise, and it is reasonable to think that the mandatory operational rules significantly affected the designers' perceptions in this study. Additionally, the impact of AI on work efficiency was a major theme for designers, likely related to the strong pressure on productivity. However, although this study acknowledges these practical constraints, we believe that it accurately captures the challenges that designers face in situations that are common in such workplaces.

Furthermore, our workplace case captured long-term user interaction with the same AI system. For example, many designers initially struggled but adapted their workflows and fundamental creative thinking process to collaborate with AI tools. This adaptation illustrates collaboration potential often missed in lab settings, prompting reflection on conventional creative processes and designers' evolving roles. Future designs and operations should facilitate such changes.

However, it is noteworthy that not all designers could embody effective collaboration. Designers who clearly divided roles between AI and themselves (e.g., P2, P6) viewed effectiveness and design as somewhat independent, trusting AI more for effectiveness but believing human adjustments could harmonize the final design. These designers acknowledged the uncertainty of AI predictions but valued the objective metrics based on extensive data. Conversely, designers who saw design and effectiveness as inseparable (e.g., P5) distrusted AI and struggled to find ways to collaborate, experiencing conflict.
These findings on individual differences align with prior research on trust and collaboration with AI outside the creative field. For instance, Lee and See \cite{Lee2004-qg} emphasized that goal-related expectations about AI's role and capabilities are crucial for defining trust. P5's comment, \textit{``Personally, I think [work efficiency has decreased]. But clearly, for inexperienced people, it has increased, so it's good for the company overall,''} highlights the role of experience. This aligns with previous findings that perceptions of self-capabilities and expertise may moderate beliefs about AI's abilities to perform a specific task \cite{Glikson2020-kw, Logg2019-bd}, and experts used AI-provided advice less than lay participants did \cite{Logg2019-bd}. This suggests that higher self-confidence in one's abilities might reduce trust in AI.

Furthermore, the contrast between P5 and P6 highlights a fundamental philosophical divergence in their approaches to design. P6 remarked that subjective judgment might not be appropriate when, for example, a woman in her 20s is creating advertisements for men in their 60s, whereas P5 argued, \textit{``Ads are seen by humans. So, if the human creating it doesn't find it interesting, it won't resonate with others.''} These contrasting views demonstrate that trust in AI depends on design philosophy and self-confidence in one's abilities. The key is not to debate which perspective is correct but to explore and design AI interactions that enable better collaboration regardless of the designer's viewpoint. Pinski et al. \cite{Pinski2023-sk} found that humans leveraging AI knowledge closely align their delegation decisions with task suitability, improving task performance. This implies that knowledge-based appropriate delegation fosters better collaboration and performance. Thus, providing appropriate knowledge to users may be one of the future recommendations to promote suitable expectations, trust, and delegation, enhancing performance.

\subsection{Design Recommendations}
Finally, we offer the following design recommendations for effective collaboration between designers and decision-making AI in the creative design industry:
1. Instruct AI capabilities,
2. Implement XAI features,
3. Ensure clear role delegation.

Firstly, ensuring users understand AI's technical strengths and fostering appropriate expectations of capabilities can promote effective collaboration. For decision-making AI, emphasizing its objective judgments based on extensive data is crucial. As users are sensitive to prediction uncertainty and may fluctuate their trust in every prediction/result error in a myopic view, leading them to care more about the long-term overall effect, e.g., presenting average performance, can help maintain stable and appropriate trust, eventually supporting appropriate reliance on AI.

Secondly, designers in interviews highlighted the need for XAI from multiple perspectives: trust, efficiency, and individual skill growth. In fact, XAI has been shown to induce appropriate reliance by improving transparency and interpretability \cite{Scharowski2022-zo, Schemmer2023-xu}. In particular, when integrating decision-making AI into creative workflows, it is necessary for users to alternately repeat the process of accessing the model to check the score and creating the design to improve the score. These characteristics are increasing the demand for XAI. Specifically, providing feedback on the reasons why a score is high or low either visually (e.g., Grad-CAM \cite{Selvaraju2017-ux, Sawada2024-ws} and LIME \cite{Ribeiro2016-dn, Sawada2024-ws}) or by text (e.g., natural language feedback \cite{Wu2024-jn, Duan2024-zb}) would help users better understand the model's behavior, easily find ways to improve the scores, and learn elements of effective design.

Lastly, our study suggested that appropriate delegation and assignment of roles, based on designers' trust in AI, help overcome conflicts between human and AI judgments. As previous studies on decision-making AI have shown that designing appropriate task delegation between AI and humans can contribute to effective human-AI collaboration \cite{Lai2022-ey, Pinski2023-sk}, in the creative domains as well, it would help human designers effectively incorporate both subjective and objective aspects into their designs.
This may specifically include an agentic approach, such that an agent that enables non-linear collaboration in creative design \cite{Zhou2024-fd}.
It should also be considered that appropriate roles and task assignments may change depending on the users and the situation. As Ashktorab et al. \cite{Ashktorab2024-ro} pointed out, the nature of human-AI collaboration is increasingly dynamic due to LLMs, leading to \textit{``co-adaptation''} that requires flexible role adjustment by both AI and users.

\subsection{Limitations and Future Work}
We have several limitations in our study, prompting future studies to complement our findings.
Firstly, there may be inherent biases among the group that the informant belong to. For instance, designers with strong negative views on AI do not work at the company or might have left if they couldn't overcome conflicts, leading to a potential bias unique to the group. Thus, our insights might reflect a more optimistic view of AI integration, potentially overlooking challenges for those who do not have an affinity with emerging technologies.
Therefore, our findings might offer a skewed understanding of AI's overall impact on design work, emphasizing the need for further controlled quantitative research.

Furthermore, the non-independent relationship between some of the authors and the case organization presents a methodological limitation of this study.
Specifically, structural power dynamics, namely, the relationship between interviewers from the parent company and interviewees from the subsidiary, may have discouraged participants from sharing negative experiences and contributed to socially desirable responses, although we emphasized to interviewees that their participation was voluntary, their responses would remain anonymous, and that there would be no negative consequences for expressing critical views toward the system or organization.
Additionally, most of the authors, including the interviewers, were employees of the company that developed AdScoreAI. This close affiliation may have led to the internalization of organizational practices and a potential weakening of critical reflection.
While we took extensive measures, as described in the Method section, to mitigate these risks and aimed to ensure transparency and neutrality throughout the reporting process, we acknowledge that some latent bias may still be present in our findings.

Additionally, long-term negative impacts of conflicts, such as burnout, were outside the scope of this study but could threaten designers' identities in the long run, as seen in other studies on professional identity in the banking industry \cite{Strich2021-fz} and generative AI's impact on creative professionals \cite{Uusitalo2024-nx}. Future studies are needed to investigate the potential downsides that were not fully addressed in this study, but we believe that the insights we have gained on how designers overcome conflict will be beneficial in addressing these long-term issues as well.

Finally, cultural and linguistic backgrounds may have influenced our findings. This study was conducted in Japan, in the Japanese language. As cultural characteristics, creativity, design thinking, trust in AI, and adherence to workplace norms may be heavily influenced by Japanese culture and nationalities.
In fact, there exist cultural differences in workplace social norms, but it has been shown to be complex.
Akahori et al.'s study \cite{Akahori2024-km} has shown that, in spite of the fact that Japan is often characterized by a collectivistic culture with strong social norms, Americans perceive stronger norms and demonstrate a higher willingness to conform to norms compared to the Japanese. Thus, future research is encouraged to deepen this direction, further taking into account the affinity with AI and how it may confound with trust, which would complement the findings of this study.
On the other hand, a limitation of our study from a linguistic perspective is that the words for trust and reliance are often used interchangeably in Japanese. The authors translated participants' comments based on context and definitions \cite{Kahr2024-sp}, which potentially lost nuances in original remarks.

\section{Conclusion}
This study builds upon prior CSCW work on human-AI collaboration in creative design and decision-making. Our research bridges these two distinct research fields, investigating the use of decision-making AI in creative design practice. By conducting a case study, we aimed to identify opportunities and challenges regarding the introduction of decision-making AI into the creative workplace. 
Findings from interviews with 12 professional designers identified that the designers' trust in AI depends on their expectations of accuracy and role. Additionally, we identified what kinds of benefits, challenges, and conflicts the designers found, and how they cope with them.

As AI technologies continue to evolve, creative professionals are not only forced to transform their traditional workflows but also redefine the creative process itself, leading to new paradigms in design practice.
The online advertising industry is among the first to shift towards a highly performance-oriented data-driven design process due to continuous objective tracking of performance. Nonetheless, we believe that these kinds of changes are likely to occur in all industries as human activity shifts online. For example, the production of articles to increase page views, the production of sound sources and music to increase the number of views, and the annotation work to improve the quality of annotations are already thought to have a similar structure.
Furthermore, even outside the creative field, as it becomes possible to measure the effectiveness of business decisions made through communication,  such as business negotiations online, it is possible that people will be required to engage in effectiveness-driven communication. As with the changes observed in our research, this may promote effective communication among people with no experience or insufficient skills, but it may also cause conflict among those with experience. We believe that the practical implications gathered in this study can be applicable to such challenges in different industries.

\section{Acknowledgments}
AI tools, such as ChatGPT and DeepL, were used for grammar checking, translation assistance, and wording refinement.
However, we declare that we originated and created the idea and work.

\bibliographystyle{ACM-Reference-Format}
\bibliography{paperpile}



\end{document}